\documentclass[aps,prl,reprint,superscriptaddress]{revtex4-1}
\usepackage{graphicx}
\usepackage{amsmath}
\usepackage{bm}
\usepackage{dcolumn}
\usepackage{amsfonts}
\usepackage{amssymb}
\usepackage[breaklinks=true]{hyperref}
\hypersetup{colorlinks=true, citecolor=blue, filecolor=blue, linkcolor=blue, urlcolor=blue}

\begin{document}
	
\title{Comparison of the magneto-Peltier and magneto-Seebeck effects in magnetic tunnel junctions}
\author{J. Shan}
\email[]{j.shan@rug.nl}
\author{F. K. Dejene}
\author{J. C. Leutenantsmeyer}
\author{J. Flipse}
\affiliation{Physics of Nanodevices, Zernike Institute for Advanced Materials, University of Groningen, Nijenborgh 4,
	9747 AG Groningen, The Netherlands}
\author{M. M\"unzenberg}
\affiliation{Institut f\"ur Physik, Greifswald University, Felix-Hausdorff-Strasse 6, 17489 Greifswald, Germany}
\author{B. J. van Wees}
\affiliation{Physics of Nanodevices, Zernike Institute for Advanced Materials, University of Groningen, Nijenborgh 4, 9747 AG Groningen, The Netherlands}
\date{\today}

\begin{abstract}
Understanding heat generation and transport processes in a magnetic tunnel junction (MTJ) is a significant step towards improving its application in current memory devices. Recent work has experimentally demonstrated the magneto-Seebeck effect in MTJs, where the Seebeck coefficient of the junction varies as the magnetic configuration changes from a parallel (P) to an anti-parallel (AP) configuration. Here we report the study on its as-yet-unexplored reciprocal effect, the magneto-Peltier effect, where the heat flow carried by the tunneling electrons is altered by changing the magnetic configuration of the MTJ. The magneto-Peltier signal that reflects the change in the temperature difference across the junction between the P and AP configurations scales linearly with the applied current in the small bias but is greatly enhanced in the large bias regime, due to higher-order Joule heating mechanisms. By carefully extracting the linear response which reflects the magneto-Peltier effect, and comparing it with the magneto-Seebeck measurements performed on the same device, we observe results consistent with Onsager reciprocity. We estimate a magneto-Peltier coefficient of 13.4 mV in the linear regime using a  three-dimensional thermoelectric model. Our result opens up the possibility of programmable thermoelectric devices based on the Peltier effect in MTJs.
\end{abstract}
\maketitle

\begin{figure}[t]
	\includegraphics[width=8.5cm]{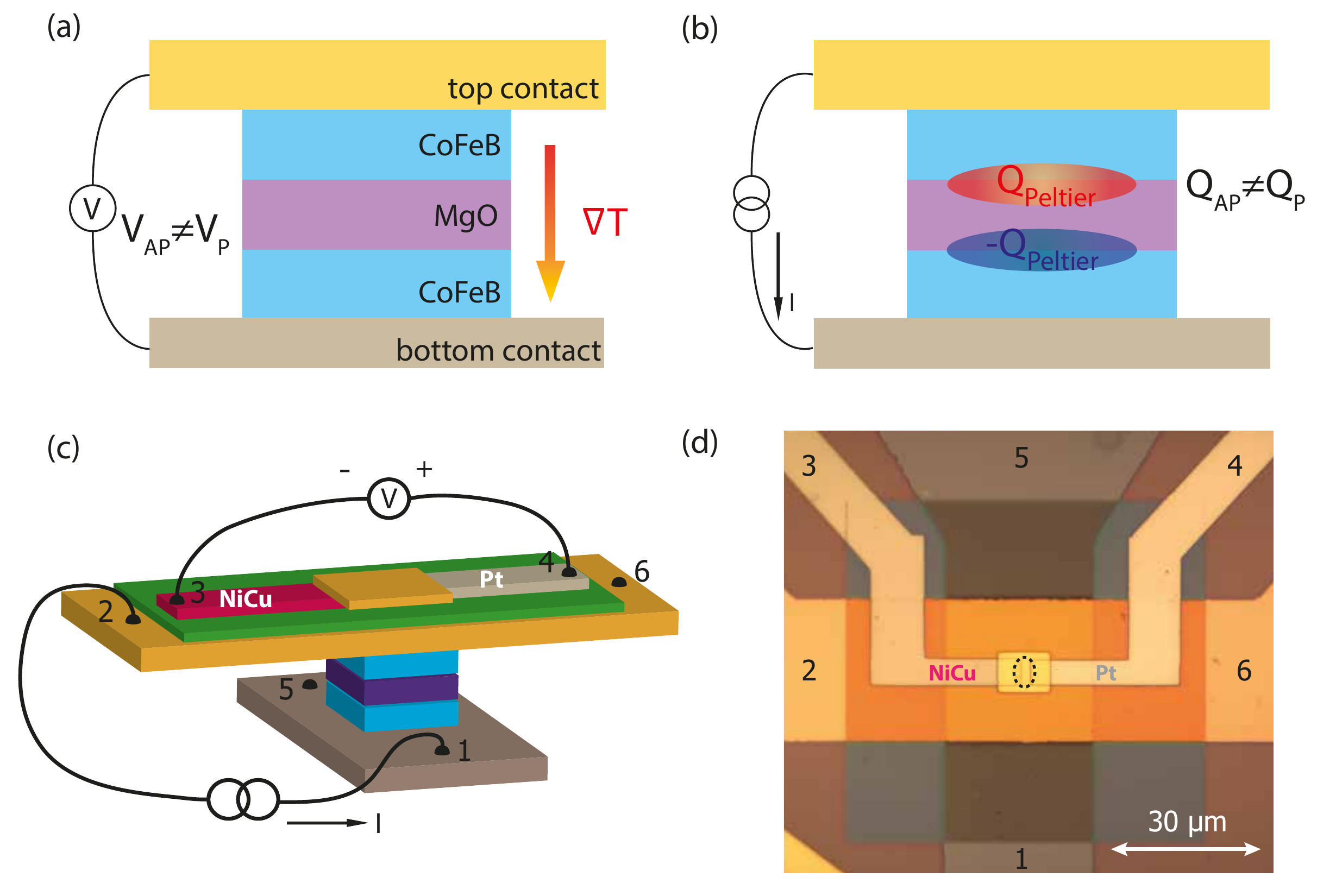}
	\caption{Concept and device geometry. (a), Concept of the magneto-Seebeck effect. A temperature gradient is applied across an MTJ, resulting in a Seebeck voltage that is dependent on the magnetic configuration. (b), Concept of the magneto-Peltier effect. A charge current is sent through an MTJ, resulting in a Peltier heating/cooling at the interfaces that also depends on the magnetic configuration. Joule heating is not shown for simplicity. (c), Schematic representation of the measured device. A Pt-NiCu (constantan) thermocouple that is electrically isolated from the top contact by an Al$_2$O$_3$ layer (green color) is used to detect temperature changes. In the Peltier measurement, charge current is sent through the pillar (from contact 1 to 2), while recording the voltage over the thermocouple (contact 3 and 4), as plotted here. Contacts 5 and 6 are used for 4-probe TMR measurements. For the reciprocal Seebeck measurement, current is sent through the thermocouple while recording the voltage over the pillar using contacts 1 and 2. (d), Optical microscope image of the measured device. The dotted circle indicates the location of the MTJ. The size of the junction measured in the main text is 2.7$\times$5 $\mu$m$^2$ by size.}
	\label{figure01}
\end{figure}

The electrical resistance of a magnetic tunnel junction (MTJ), a stack of two ferromagnetic layers separated by an insulating tunnel barrier, depends on the relative magnetic orientation of the two magnetic layers \cite{miyazaki_giant_1995,parkin_giant_2004,yuasa_giant_2004}. This tunnel magnetoresistance (TMR) effect puts MTJs at the forefront of the applications in the field of spintronics \cite{zutic_spintronics:_2004}. Spin caloritronics \cite{bauer_spin_2010,bauer_spin_2012,boona_spin_2014} is an emerging field that couples thermoelectric effects with spintronics. Many interesting physical phenomena were discovered such as the spin (-dependent) Seebeck effect in magnetic metal \cite{slachter_thermally_2010}, magnetic semiconductor \cite{jaworski_observation_2010} and magnetic insulator \cite{uchida_spin_2010}. Particularly, in spin tunneling devices, the magneto-Seebeck effect was theoretically studied \cite{czerner_spin_2011,heiliger_ab_2013,wang_thermoelectricity_2014,lopez-monis_tunneling_2014} and experimentally observed \cite{walter_seebeck_2011,liebing_tunneling_2011,lin_giant_2012,zhang_seebeck_2012,teixeira_giant_2013} in MTJs, where the Seebeck coefficient of the junction can be varied by changing the magnetic configuration. More recently, the spin (-dependent) Peltier effect that is driven by spin (polarized) currents has been experimentally observed in metallic \cite{gravier_spin-dependent_2006,flipse_direct_2012} and insulating ferromagnets \cite{flipse_observation_2014}, which are shown to obey the Thomson-Onsager reciprocity relation \cite{onsager_reciprocal_1931,dejene_verification_2014,avery_peltier_2013} to the spin (-dependent) Seebeck effect. From this relation, the reciprocal effect of the magneto-Seebeck effect, which can be named as magneto-Peltier effect, is also expected in MTJs (see Fig.~\ref{figure01}(a)(b)).  

However, experimental studies of the magneto-Peltier effect have not been reported so far. Its small effect compared to the often-dominant Joule heating effects has left the experimental observation elusive. In this work, we report the first experimental study of the magneto-Peltier effect as well as higher order heating effects, and compare the Peltier measurements to the Seebeck measurements on the same junction. Via sensitive thermometry architecture and measurement techniques, we are able to measure small temperature changes as well as distinguish the linear (due to Peltier effect) and nonlinear effects (due to Joule heating). 

Although both the electric conductance and Seebeck coefficient depend on the relative magnetic configuration in a MTJ, the mechanisms behind them are not the same. While the electric conductance is determined by the transmission probability $T_{P,AP}(E)$ of electrons across the insulating barrier around the Fermi energy $E_F$, the Seebeck coefficient $S_{P,AP}$ solely depends on the electron-hole asymmetry of $T_{P,AP}(E)$ around $E_F$. By Onsager reciprocity, the Peltier coefficient $\Pi$ is closely related to $S$ by $\Pi=ST_0$, where $T_0$ denotes certain temperature. Using the expression for $S$ \cite{czerner_spin_2011,walter_seebeck_2011} we can express $\Pi$ as:
\begin{equation}
	\Pi _{P,AP}=-\frac{\int T_{P,AP}(E)(E-E_{F})(-\partial _{E}f_{0})dE}{e\int T_{P,AP}(E)(-\partial _{E}f_{0})dE},
	\label{eq:Pi}
\end{equation}
where $e$ is the elementary charge and $f_0$ is the Fermi-Dirac distribution function at temperature $T_0$. For a constant tunneling current $I$ through the MTJ, the Peltier heat $Q_{\Pi}$ carried by this current is different for the P and AP configurations, leading to different temperature biases across the junction between the two configurations. Disregarding Joule heating and assuming that the thermal conductance $\kappa$ of the MTJ is independent of the magnetic configuration, the temperature difference between the parallel and antiparallel configuration $\Delta T=\Delta T_{AP}-\Delta T_{P}$ can be estimated by balancing the Peltier heat $Q_{\Pi}$ with the backflow of the heat current through the junction as
\begin{equation}
\Delta T=\frac{tI}{{\kappa}A}(\Pi_{AP}-\Pi_{P}),
\label{eq:deltaT}
\end{equation}
where $t$ is the thickness of the tunnel barrier and $A$ is the area of the junction. Using the parameters given in ref. \cite{walter_seebeck_2011}, it can be estimated that for an electric current density of 5$\times$10$^3$ A/cm$^2$ through the junction, the change in temperature due to the magneto-Peltier effect can reach $\sim$100 $\mu$K at room temperature, which requires sensitive thermometry technique, as we use here. 

Fig.~\ref{figure01}(c)(d) show the device geometry and measurement configuration as employed in our experiment. We study the CoFeB/MgO/CoFeB MTJs, same as in ref. \cite{walter_seebeck_2011} that are reported to have a large magneto-Seebeck effect \cite{walter_seebeck_2011,liebing_tunneling_2011,czerner_spin_2011} due to their half-metallic transmission property \cite{yuasa_giant_2004}. The layer structure of the patterned MTJ sample stack (from bottom to top) is: Si (100)/SiO$_2$ 500 nm/Ta 15 nm/CoFeB 2.5 nm/MgO 2.1 nm/Co-Fe-B 5.4 nm/Ta 5 nm/Ru 3 nm/Cr 5 nm/Au 25 nm. Detailed fabrication processes can be found in ref. \cite{walter_seebeck_2011}. To sense the temperature change locally at the top of the junction, a thermocouple consisting of constantan (Ni$_{45}$Cu$_{55}$) and Pt is fabricated over the top contact of the MTJ \cite{flipse_direct_2012,dejene_spin_2013,dejene_verification_2014}. Constantan is dc sputtered and Pt is e-beam evaporated, both 90 nm in thickness. A 50-nm-thick Al$_2$O$_3$ layer is e-beam evaporated over the top contact to electrically isolate MTJ and thermocouple from the top contact, so that no charge-related effects are picked up by the thermocouple. Finally, a 130 nm layer of Au is deposited to connect the two arms of thermocouple, creating a uniform temperature distribution over the junction. In the Peltier measurement configuration, we send a charge current through the pillar (contact 1 to 2) while recording the thermovoltage using the thermocouple (contact 3 to 4). Note that in this configuration the temperature changes resulting from both the Peltier and Joule heating effects are measured. Meanwhile we also record the tunnel magnetoresistance (TMR) by four-probe method using contacts 5 and 6. All measurements shown in the main text are performed on a single device at room temperature. Measurements on two other samples can be found in the supplementary.

A standard lock-in technique is used for our measurements, where an ac current is sent through the system at a low excitation frequency ($\sim$17 Hz), so that a steady-state temperature condition is reached and at the same time capacitive coupling is suppressed. The voltage output from the sample is separated into different harmonic signals ($V^{1f},\ldots V^{nf}$) in terms of the input frequency. With this technique it is possible to isolate the Peltier effect that is linear with current from the Joule heating effect that is of second or even higher responses. In a simple case where only $V^{1f}$ and $V^{2f}$ are present, Peltier effect is linked to $V^{1f}$ while the secondary Joule heating effect is linked to $V^{2f}$. But in a more complex case where higher order effects are also present, the $n$-th order signal is not directly equal to the $n$-th harmonic signal; instead, all higher harmonic signals (with the same parity as $n$) that are nonzero need to be included by a straightforward algebraic operation (see supplementary for more details). 
\begin{figure}[t]
	\includegraphics[width=8.5cm]{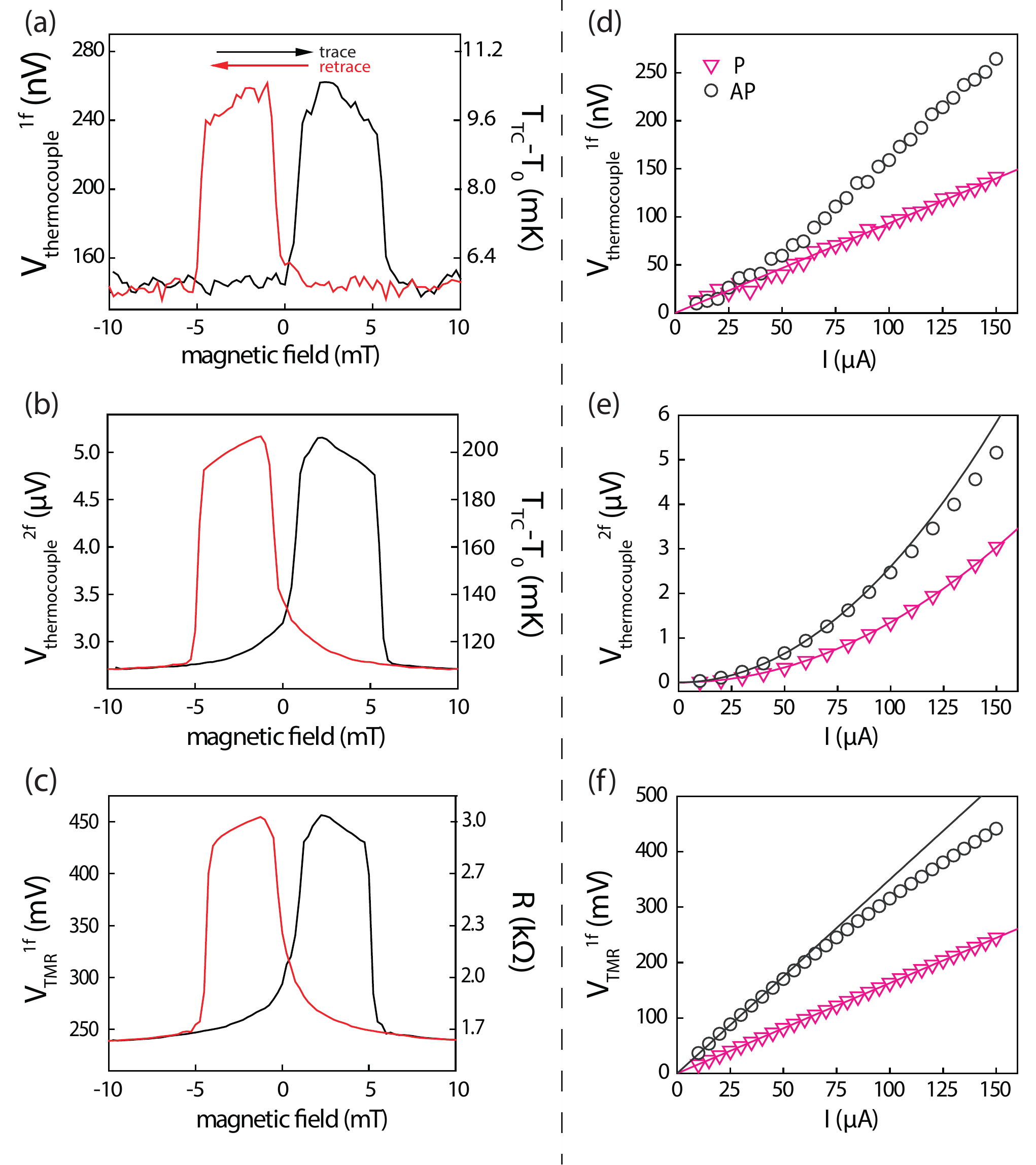}
	\caption{Lock-in signals from magneto-Peltier and TMR measurement configurations. (a)(b), First ($V^{1f}$) and second ($V^{2f}$) harmonic signal measured at the thermocouple for an r.m.s current of 150 $\mu$A. c, First ($V^{1f}$) harmonic signal of TMR measured at the same current. On the right axes the temperature differences detected by the thermocouple relative to the room temperature $T_0$ are given as $T_\textup{TC}-T_{0}=V^{1f(2f)}/(S_{\textup{Pt}}-S_{\textup{NiCu}})$, where $T_0$ is 290 K. (d)-(f), Current dependence of the corresponding measurements, for P (open triangles) and AP (open circles) configurations, where P (AP) is obtained when setting the magnetic field at 10 (2) mT. Solid lines are linear (in d, f) or quadratic fits (in e), as references.}
	\label{figure02}
\end{figure}

Fig.~\ref{figure02} shows the experimental results of the magneto-Peltier measurements. We apply an ac current of 150 $\mu$A (r.m.s) through the MTJ while sweeping the magnetic field. Both the first and second-harmonic voltages recorded at the thermocouple ($V^{1f}$ and $V^{2f}$), shown in Fig.~\ref{figure02}(a)(b), exhibit four abrupt changes corresponding to the switching from P to AP configuration and back, implying a change in the temperature at the top contact.  $V^{1f}$ of TMR is shown in Fig.~\ref{figure02}(c). We measure the current dependence of these signals correspondingly, for both the P and AP configurations, as shown in Fig.~\ref{figure02}(d)-(f). We did not apply ac currents higher than 150 $\mu$A to avoid dielectric breakdown of the MTJ ($\sim$2 V across the junction for 2.1 nm MgO) \cite{khan_dielectric_2008}. In the P configuration we have a simple case where no higher harmonic signals can be detected, and the $V^{1f}$ ($V^{2f}$) signal detected at the thermocouple is linear (quadratic) with the current, which can be considered as the Peltier signal (Joule heating signal). However, for AP configuration, we have a more complex case where higher order effects are present. The $I-V$ characteristic of the MTJ (see supplementary) is nonlinear with the current, or in other words, the resistance of the junction $R$ is bias-dependent \cite{moodera_large_1995,zhang_quenching_1997,valenzuela_spin_2005}. Therefore, the Joule heating effect ($I\cdot V$) is not only present in the second order, but also brings on higher order responses. The consequence of this nonlinearity is twofold: first, $R$ decreases with both larger positive and negative biases. This leads to an even higher order responses at the thermocouple which deviate the $V^{2f}$ from a quadratic behavior, as shown in Fig.~\ref{figure02}(e). In addition, $R$ also shows an asymmetric dependence for $+I$ and $-I$, i.e., $V(+I) \neq V(-I)$, indicating that the dissipation at the junction is different when the bias is reversed. The reason for this asymmetry can be attributed to the inevitable difference between the two interfaces across the MgO \cite{valenzuela_spin_2005}. Although this effect is only present in higher odd order heating signals on thermocouple, it mimics a Peltier-like effect and strongly deviates $V^{1f}$ from a linear behavior (see Fig.~\ref{figure02}(d)). A more extensive quantitative analysis can be found in the supplementary.

\begin{figure*}[t]
	\includegraphics[width=18cm]{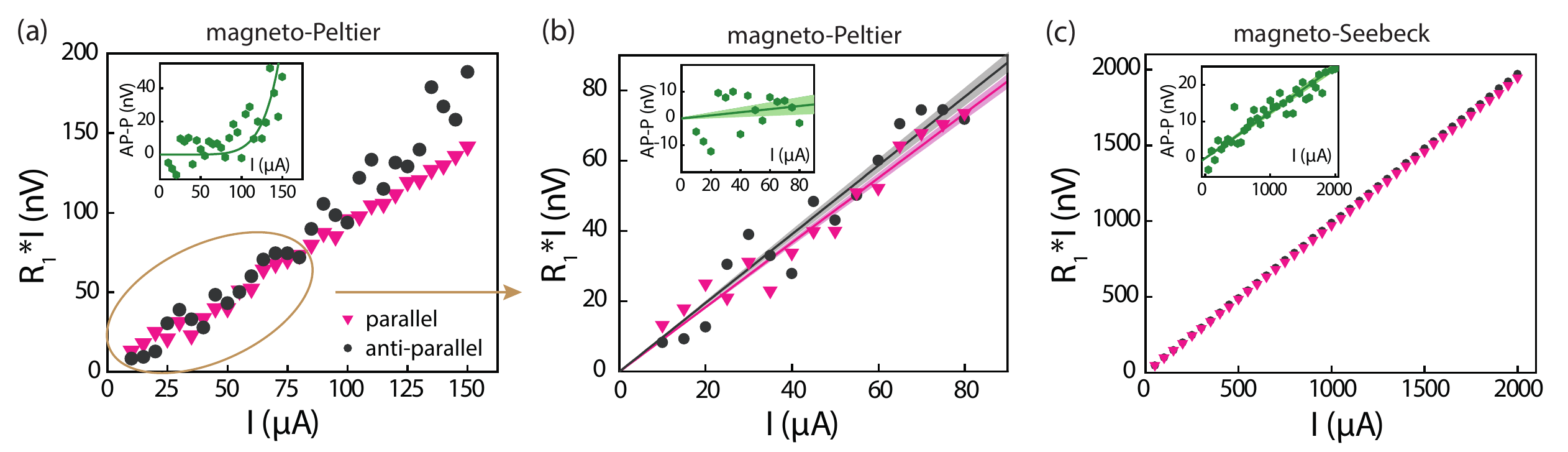}
	\caption{Comparison of the magneto-Peltier and magneto-Seebeck linear responses. (a), The extracted linear response calculated up to including the $V^{5f}$, as a function of applied current. The circled part is where the $V^{7f}$ is close to 0 and is therefore represents pure linear signals. Inset shows the difference between the P and AP configurations which is fitted to a seventh order behavior. (b), Linear fitting of the circled part in (a). The shaded zones indicate the standard deviations of the fitted slopes. Inset shows the difference between the P and AP configurations which is linearly fitted, with the shaded area indicating one standard deviation. (c), Current dependence of the Seebeck measurements, without any compensations from higher harmonic signals. Inset shows the difference between the AP and P configurations which is fitted to a linear behavior. }
	\label{figure03}
\end{figure*}

It is therefore important to determine the pure linear signal ($R_1I_0$) of the AP case in order to discuss the magneto-Peltier effect. Taking advantage of the lock-in detection technique, we can measure the higher harmonic signals ($V^{3f}, V^{5f}\ldots$) by tuning the lock-in detection frequency to the corresponding harmonics frequency. The first order (linear) response can be expressed as a linear combination of the odd harmonics voltages as $R_1I_0=V^{1f}+3V^{3f}+5V^{5f}\ldots$ (see supplementary). Here we only include higher harmonic signals up to $V^{5f}$, as $V^{7f}$ cannot be determined accurately within our noise level ($\le\left | 5\ \textup{nV} \right |$). The results are shown in Fig.~\ref{figure03}(a). However, the difference between P and AP still shows a seventh order behavior as a function of the current. This means that, especially at larger currents (above 80 $\mu$A), the seventh order response is still present; however, the  $V^{7f}$ signal that can be measured is only 1/8 of the seventh order response, which made it difficult to be included to extract $R_1I_0$. Nevertheless we can still rely on the lower-current regime before the onset of the seventh order response (below 80 $\mu$A, circled part in Fig.~\ref{figure03}(a)), which can be regarded as purely linear regime. We fit the curves for P and AP individually for this regime and especially focus on their difference, which can be considered as the magneto-Peltier effect. Although the difference is small compared to the noise, we fit it linearly and estimate a slope range bounded by one standard deviation, 58$\pm$35 $\mu \Omega$, as shown in the inset of Fig.~\ref{figure03}(b).  

To support our estimation of the magneto-Peltier signal, we perform magneto-Seebeck measurements on the same device by effectively reversing the role of the current and voltage contacts as used in the Peltier measurement. Here we send an ac current through the thermocouple (contact 3 to 4) thereby creating a vertical temperature gradient over the MTJ via the Peltier heating/cooling ($\propto I$) at the NiCu-Au and Au-Pt interfaces. The Seebeck voltage (open-circuit thermovoltage) due to this vertical temperature gradient is measured using contacts 1 to 2. In Seebeck measurements, it is possible to send larger currents up to 2 mA through the thermocouple with a resistance of 190 $\Omega$. Unlike the Peltier measurement, in the Seebeck measurement no higher odd harmonic features are observed for either P or AP configurations, implying a linear behavior for the Seebeck signal in the measured current range. This is because the thermocouple is purely ohmic, in contrast to the nonlinear MTJ. The current dependence of the magneto-Seebeck measurements are shown in Fig.~\ref{figure03}(c). According to the Thomson-Onsager reciprocity relation, the linear response signals for the Peltier and Seebeck effect should be the same, as well as the difference between P and AP configurations \cite{dejene_verification_2014}. From Fig.~\ref{figure03}(c), the difference between the two configurations is 12.5$\pm$0.4 $\mu \Omega$, which falls into the estimated range of the magneto-Peltier effect within 2 standard deviations (corresponds to a confidence level of 95\%), therefore showing no statistically significant difference. This is consistent with the reciprocity between the magneto-Seebeck and magneto-Peltier measurements. In our opinion, there is no fundamental reason for the rather large difference in the average values for magneto-Peltier and Seebeck coefficients, except for the experimental difficulties in obtaining the magneto-Peltier coefficient. Note that the background signals for Seebeck and Peltier configurations correspond closely, indicating the validity of our approach. However, the backgrounds contain Seebeck/Peltier effects from all metal interfaces, and therefore are not directly linked to the Seebeck/Peltier coefficients of the MTJ. 

By using a three-dimensional finite element model (3D-FEM) \cite{slachter_modeling_2011} we can quantify our results. We focus on the estimation of the relative change of the Peltier coefficient from P to AP configuration of the MTJ. We do not model the electron tunneling process, but regard MgO as a conductor whose electrical conductivity and Peltier coefficient vary between P and AP states, while keeping other properties of the MTJ constant. The details can be found in supplementary. We find that the modeled magneto-Peltier signal is very sensitive to the choice of the thermal conductivity of the MgO layer ($\kappa_{\textup{MgO}}$) and the difficulty of measuring this quantity directly can create a big uncertainty in our estimation. Here we adopt the same value from ref. \cite{walter_seebeck_2011}, where $\kappa_{\textup{MgO}}$ = 4 W/(m$\cdot$K) was used for 2.1 nm MgO layer, taking into account both the crystalline quality of MgO and its thermal interfaces with CoFeB. By fitting to our experimental result 12.5$\pm$0.4 $\mu \Omega$ from the Seebeck measurements (which has less statistical uncertainty), we obtain the change of Peltier coefficient of MgO to be $\Delta\Pi =\Pi_{AP}-\Pi_{P}=\Delta S\cdot T$ =13.4 mV from P to AP, which by Onsager relation corresponds to $\Delta S=S_{AP}-S_{P}$  =46.2 $\mu$V/K. This is close to the Seebeck coefficient change of MgO reported in ref. \cite{walter_seebeck_2011}.

In conclusion, we have observed the magneto-Peltier effect in magnetic tunnel junctions and confirmed its reciprocity to the magneto-Seebeck effect by measuring both effects in a single device. We also observed higher order heating effects which greatly enhance the magneto-Peltier signal in the large-bias regime. We attribute this effect to the asymmetric resistance of MTJ for opposite bias. In addition to providing additional insight in the nature of heat dissipation in MTJs, our results open up the possibility of magnetically controllable cooling mechanism in MTJs, which can be potentially applied in novel magnetic logic devices. 

\vspace{1cm}
\begin{acknowledgments}
The authors thank G. Reiss and T. Kuschel for helpful discussion, M. Walter for MTJ growth parameter optimization, and M. de Roosz, H. Adema, T. Schouten and J.G. Holstein for technical assistance. This work is part of the research program of the Foundation for Fundamental Research on Matter (FOM) and is supported by NanoLab NL, the priority program DFG SpinCaT 1538 within MU 1780/8-1,2, EU-FET Grant
InSpin 612759 and the Zernike Institute for Advanced Materials.

J.S and F.K.D contributed equally to this work.
\end{acknowledgments}

%
%\bibliography{references}
\onecolumngrid
\clearpage

\subsection{\Large  \uppercase{Supplemental Material}}
\subsection{\large {I. Results for two other samples}}
\begin{figure}[h]
	\includegraphics[width=15cm]{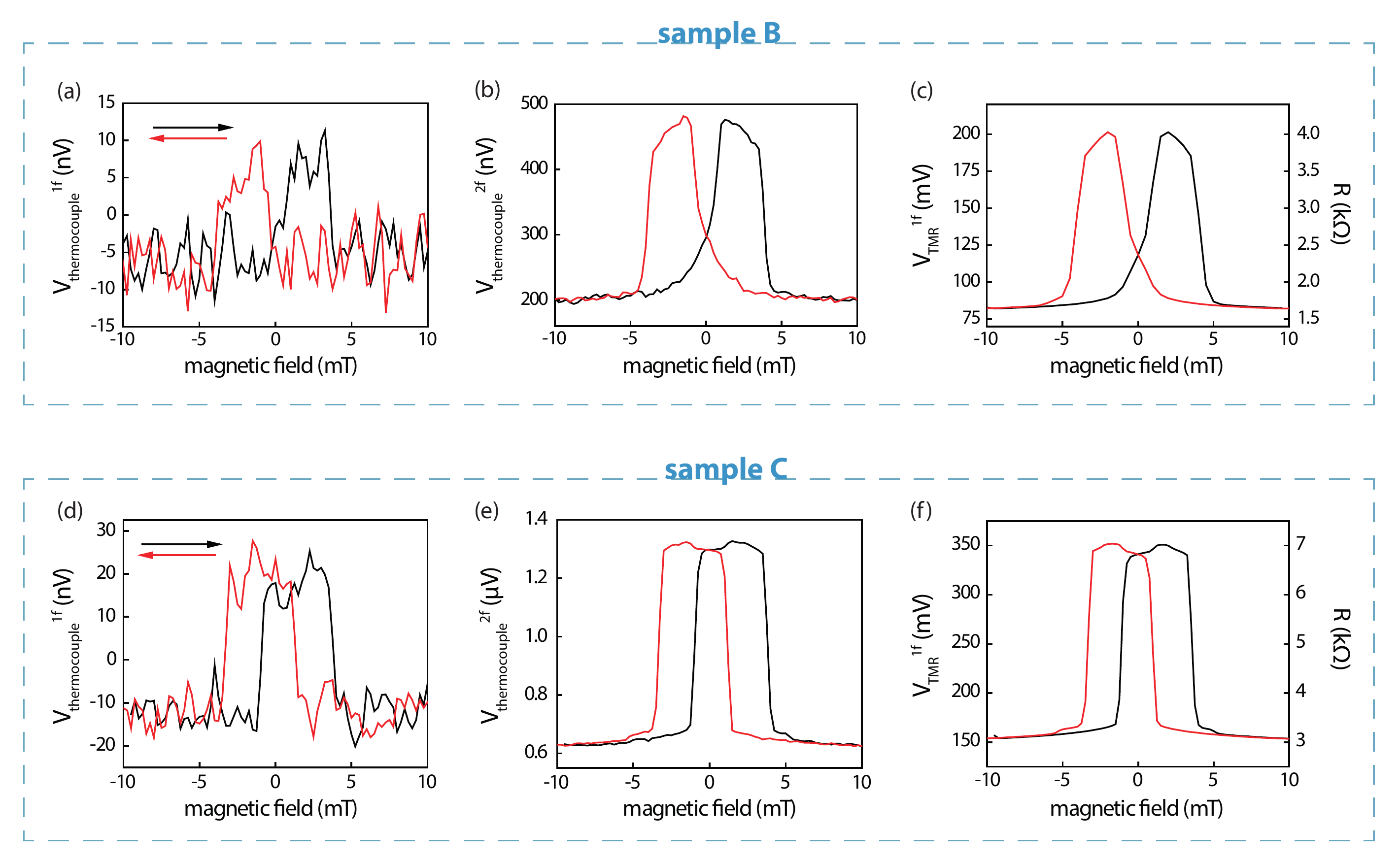}
	\caption{Magneto-Peltier and TMR measurements for two other samples. Figure (a)(b)(c) are measured on a sample where the junction size is 2.1$\times$3.2 $\mu$m$^2$. Figure (d)(e)(f) are measured on a sample where the junction size is 1.8$\times$3.1$\mu$m$^2$. Here the r.m.s current for both measurements is 50 $\mu$A. }
	\label{figs1}
\end{figure}

In the main text, all the measurements shown were performed on a single device, which we here denote as sample A. Here we show additional measurements performed on two other samples, sample B (Fig.~\ref{figs1}(a)-(c)) and sample C (Fig.~\ref{figs1}(d)-(f)). From left to right the first harmonic $V^{1f}$, second harmonic signal $V^{2f}$ of magneto-Peltier measurements and $V^{1f}$ of TMR measurement are plotted. Similar features are observed compared to Fig.~\ref{figure02} in the main text. 
\subsection{\large {II. Extraction of the linear response from the lock-in measurements}}
The measurements shown in this paper are all performed using a lock-in detection technique. In this technique, an ac current $I(t)=\sqrt{2}I_{0}\sin(\omega t)$ with an angular frequency $\omega$, generated by one of the lock-in amplifiers, is sent to the studied system as an input current. In general, the system output in terms of voltage can be written as:
\begin{equation}
V(t)=R{_{1}}I(t)+R{_{2}}I^{2}(t)+R{_{3}}I^{3}(t)+R{_{4}}I^{4}(t)+R{_{5}}I^{5}(t)+\ldots
\label{eq:Vt}
\end{equation}
where $R_nI^n(t)$ is the $n$-th order response and $R_n$ is the $n$-th order coefficient, which we want to determine. The lock-in technique can separate different harmonic contributions of $V(t)$ according to the orthogonality of sinusoidal functions: it multiplies $V(t)$ by a reference sine-wave signal, with an angular frequency that is an integral (denoted by $n$) multiple of $\omega$, and then averages this product over time as:
\begin{equation}
V^{nf}=\frac{\sqrt{2}}{T}\int_{0}^{T}\sin(n\omega t+\phi))V(t)dt,
\label{eq:Vnf}
\end{equation}
where we can obtain a signal $V^{nf}$ as the $n$-th harmonic signal directly from the output of the $n$-th lock-in amplifiers. In this way, different harmonic components of $V(t)$ are separated. However, the $n$-th harmonic signal $V^{nf}$ is often not equal to the $n$-th order response $R_n I^n (t)$, especially at large current biases where higher order responses are non-negligible, although it is possible to express one from the other. Assuming we have a voltage response only up to the fifth order, different harmonic signals can be expressed as:
\begin{subequations}
	\begin{eqnarray}
	&& V^{1f}=R_1I_0+\frac{3}{2}R_1I_0^3+\frac{5}{2}R_5I_0^5 \ \ \ (\phi=0^{\circ}), \label{eq:V1f}\\
	&& V^{2f}=\frac{1}{\sqrt2}(R_2I_0^2+2R_4I_0^4)\ \ \ (\phi=-90^{\circ}), \label{eq:V2f}\\ 
	&& V^{3f}=-\frac{1}{4}(2R_3I_0^3+5R_5I_0^5)\ \ \ (\phi=0^{\circ}), \label{eq:V3f}\\ 
	&& V^{4f}=-\frac{\sqrt2}{4}R_4I_0^4\ \ \ (\phi=-90^{\circ}), \label{eq:V4f}\\
	&& V^{5f}=\frac{1}{4}R_5I_0^5\ \ \ (\phi=0^{\circ}). \label{eq:V5f}
	\end{eqnarray}
\end{subequations}

It is then possible to find the $n$-th order response as a linear combination of $V^{nf}$, for example, for linear response, we have
\begin{equation}
R_1I_0=V^{1f}+3V^{3f}+5V^{5f}.
\label{eq:R1I0}
\end{equation}

In the magneto-Peltier measurement, we are interested in $R_1 I_0$, which reflects the (magneto-)Peltier effect. By using three lock-in amplifiers to record $V^{1f}$, $V^{3f}$ and $V^{5f}$ simultaneously, it is possible to determine $R_1 I_0$ using Eq.~\eqref{eq:R1I0}. Such calculation has been performed to obtain Fig.~\ref{figure03}(a) of the main text.

\begin{figure}[t]
	\includegraphics[width=12cm]{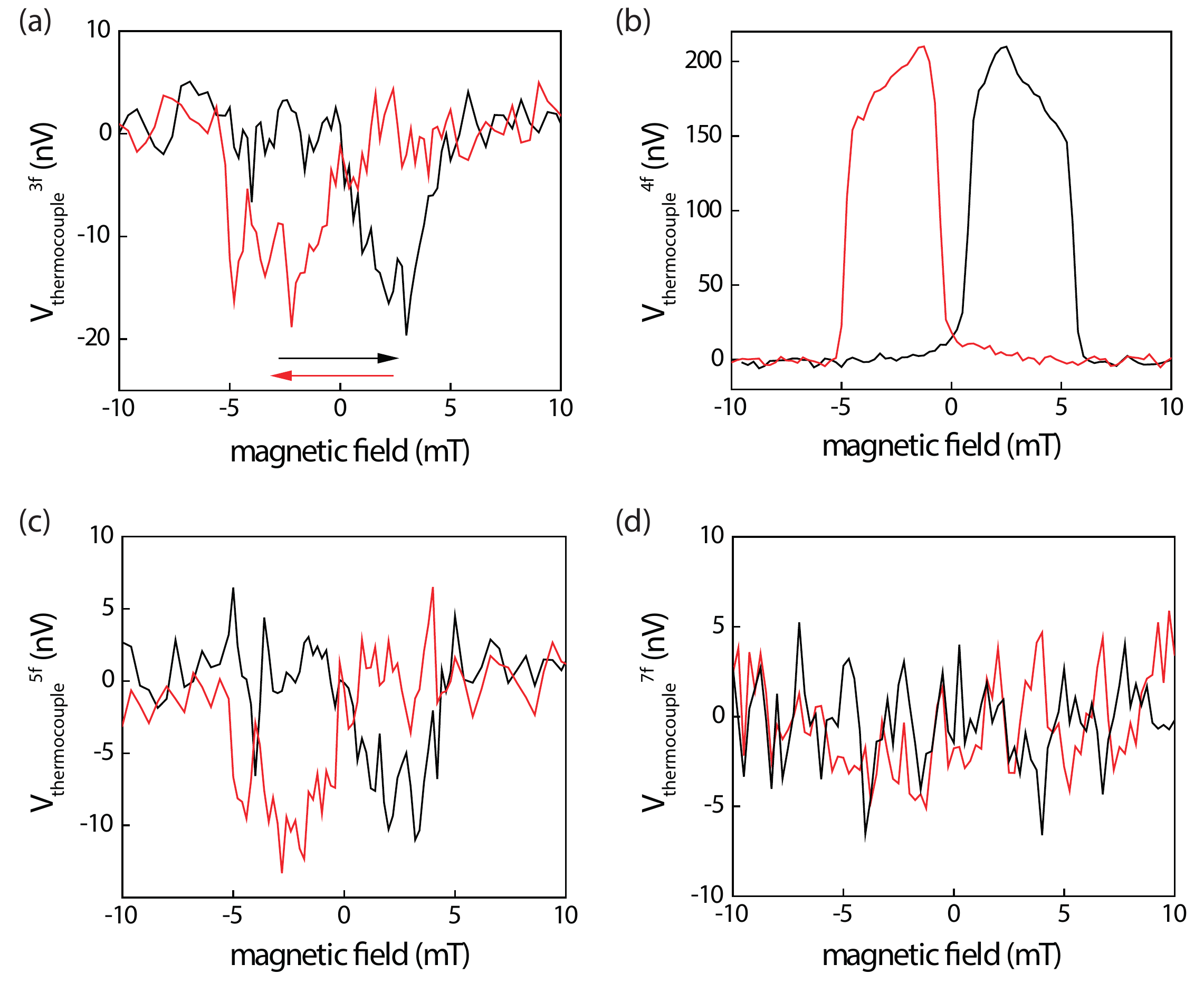}
	\caption{Higher harmonic signals for magneto-Peltier measurement. Magnetic field sweep with Peltier measurement configuration for an r.m.s current of 150 $\mu$A, where (a)$V^{3f}$, (b)$V^{4f}$, (c)$V^{5f}$ and (d)$V^{7f}$ are shown.}
	\label{figs2}
\end{figure}

Fig.~\ref{figs2} shows different higher harmonic signals $V^{3f}$, $V^{4f}$, $V^{5f}$ and $V^{7f}$, measured with an r.m.s current of 150 $\mu$A on the same sample as main text, where the higher harmonic signals are the most prominent in our measured current range. For the odd harmonic signals, switches from P to AP are visible in $V^{3f}$ and $V^{5f}$, but not in $V^{7f}$, as the signal is already within the noise level. Therefore, it is reasonable to perform Eq.~\eqref{eq:R1I0} to obtain linear response in our results. 

For the even harmonic signals, they are only relevant if we want to obtain even order responses. Here we present the magnetic field dependence of $V^{4f}$ as shown in Fig.~\ref{figs2}(b). Taking the contribution of $V^{4f}$ into consideration by using Eq.~\eqref{eq:V2f}, we can get the second order response $R_2 I_0^2$ whose dependence on the applied current is close to a quadratic behavior. 

\subsection{\large {III. Finite element thermoelectric model}}
To understand the thermoelectric transport properties and extract the Peltier coefficient of the MTJ we make use of a finite element model \cite{slachter_modeling_2011} that solves the steady-state three-dimensional current and heat equation. In this model the charge current density $\boldsymbol{\mathit{J}}$, driven by the applied voltage, and the heat current $\boldsymbol{\mathit{Q}}$, driven by the thermal gradient, are related to each other as
\begin{equation}
\begin{pmatrix} \boldsymbol{\mathit{J}} \\ \boldsymbol{\mathit{Q}} \end{pmatrix} =- \begin{pmatrix} \sigma & \sigma S \\ \sigma \Pi & \kappa \end{pmatrix} \begin{pmatrix} \boldsymbol{\nabla} V \\ \boldsymbol{\nabla} T \end{pmatrix}
\label{eq:jq}
\end{equation}
where $\sigma$ is the electrical conductivity, $S$ is the Seebeck coefficient, $\Pi=ST_0$ is the Peltier coefficient at any temperature $T_0$. Conservation of charge and Joule heating are included as
\begin{equation}
\boldsymbol{\nabla}\cdot \begin{pmatrix} \boldsymbol{\mathit{J}} \\ \boldsymbol{\mathit{Q}} \end{pmatrix} 
= \begin{pmatrix} 0 \\ \boldsymbol{\mathit{J}}^2/\sigma \end{pmatrix} 
= \begin{pmatrix} 0 \\ \sigma(\boldsymbol{\nabla}^2V+S^2\boldsymbol{\nabla}^2T+2S\boldsymbol{\nabla}V\cdot \boldsymbol{\nabla}T) \end{pmatrix} 
\label{eq:jqsource}
\end{equation}

The material parameters used in the FEM are shown in Table~\ref
{table1}. We do not model the tunneling process directly but consider the MgO as a material whose electrical or thermal parameters, such as the conductivity and Peltier coefficient, depend on the magnetic configuration of the MTJ, a similar approach as employed in Ref. \cite{walter_seebeck_2011,liebing_determination_2012,zhang_seebeck_2012}. From the four-probe resistance of the MTJ (TMR measurement), we obtain the conductivity for the parallel $\sigma_P$ and antiparallel configurations $\sigma_{AP}$ and assign these conductivity values to the MgO barrier. The thermal conductivity of the insulating MgO barrier is estimated to be 4 W/(m$\cdot$K), according to Walter \textit{et al.} \cite{walter_seebeck_2011}. The Peltier coefficient of MgO is then varied to fit our experimental results. 

\begin{table}[b]
	\caption{For Au, Pt and NiCu, $\sigma$ was measured in dedicated devices \cite{bakker_nanoscale_2012} and for Ta, $\sigma$ is obtained from Ref. \cite{hahn_comparative_2013}. The thermal conductivities $\kappa$ were obtained using the Wiedemann-Franz relation valid at the temperature of our experiment. The rest of the material parameters were taken from Walter \textit{et al.} \cite{walter_seebeck_2011}. For MgO, $\sigma$ is determined from TMR measurement and $S$ is varied as a fitting parameter.}
	\begin{ruledtabular}
		\begin{tabular}{c c c c}
			\renewcommand{\arraystretch}{2}
			Material & $\sigma$ & $S$ & $\kappa$ \\
			(thickness) & ($10^6$ S/m) & ($\mu$V/K) & (W/(m$\cdot$K)) \\
			\hline
			Au (130 nm) & 27 & 1.7 & 180 \\
			Pt (90 nm) & 4.8 & -5 & 37  \\
			Ni$_{45}$Cu$_{55}$ (90 nm) & 2 & -30 & 20  \\
			Al$_2$O$_3$ (50 nm) & 0 & - & 0.15  \\
			Au top contact (25 nm) & 18 & 1.7 & 120  \\
			CoFeB (5.4 nm, 2.5 nm) & 12 & -10 & 87 \\
			MgO (2.1 nm) & $\ast$ & $\ast\ast$ & 4 \\
			Ta (1.5 nm) & 0.75 & -5 & 5.3 \\
			SiO$_2$ (500 nm) & 0 & - & 1 \\
		\end{tabular}
	\end{ruledtabular}
	\label{table1}
\end{table}

\begin{figure}[t]
	\includegraphics[width=7cm]{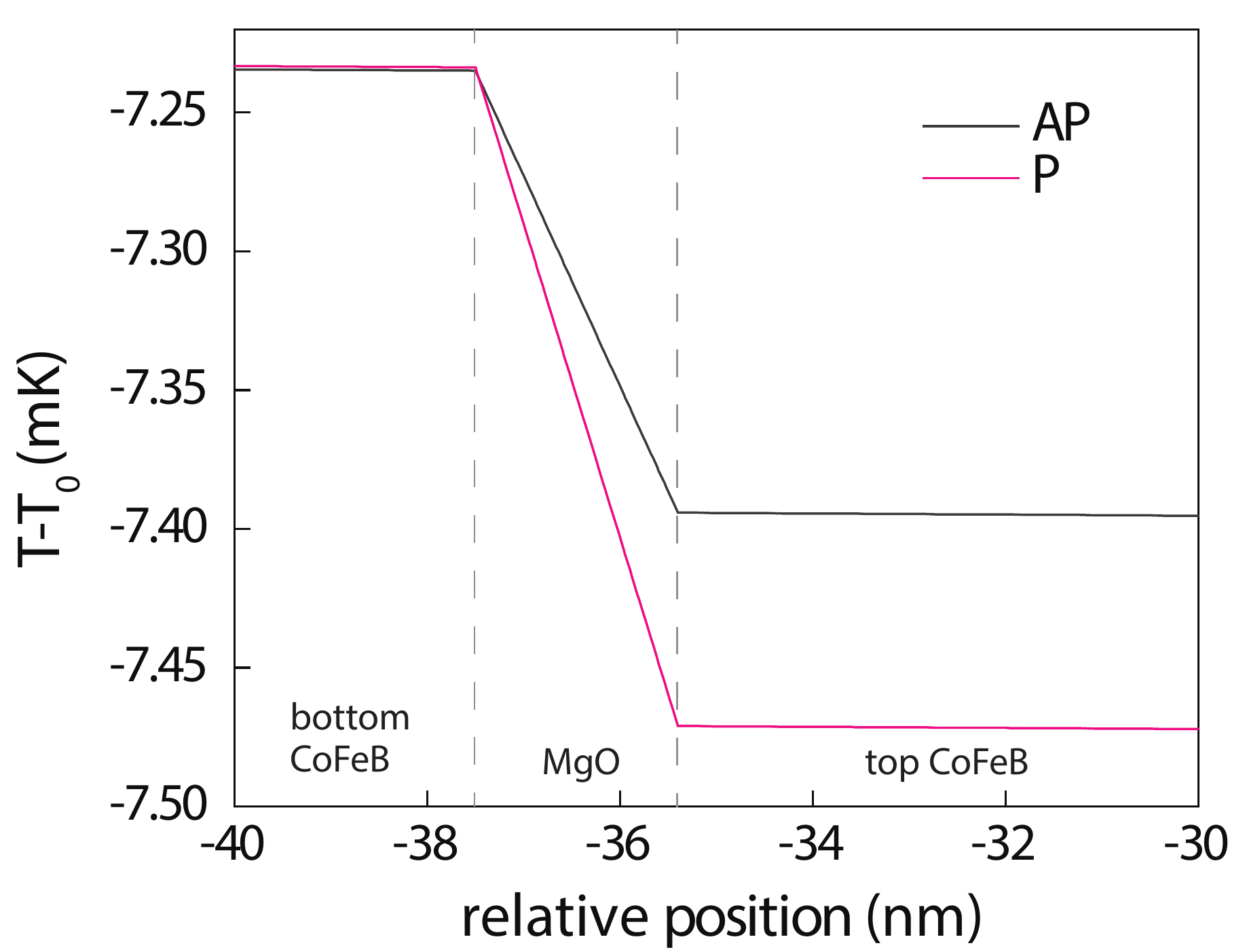}
	\caption{A finite element simulation result showing the temperature profiles due to the Peltier effect without including the dissipation. Charge current of 150 $\mu$A is sent through the MTJ resulting in a Peltier heating/cooling. If the temperature of one side of the junction; is anchored, the temperature change due to Peltier effect would be most prominent on the other side. Due to magneto-Peltier effect a different temperature at the top contact will be observed between P and AP configurations. In this figure, $S_{AP}$= -99 $\mu$V/K and $S_P$= -145.2 $\mu$V/K are used. $T_0$ is taken as 290 K.}
	\label{figs3}
\end{figure}

In the following, we describe our approach for extracting the Seebeck (Peltier) coefficient of the MTJ. We first start by obtaining the electrical conductivity of the MgO from the tunnel magnetoresistance measurements. In device A where we observed a TMR of 87.5\% (with parallel resistance $R_P$=1.6 k$\Omega$ and antiparallel resistance $R_{AP}$=3.0 k$\Omega$) when I = 150 $\mu$A, we obtain the corresponding conductivities $\sigma_P$=0.093 S/m and $\sigma_{AP}$=0.05 S/m, respectively. Although the thermal conductance of MgO is expected to depend on the magnetic configuration of the MTJ \cite{wang_thermoelectricity_2014}, for simplicity, we do not take this scenario into account. In the estimation of the magneto-Peltier coefficient, we match the measured Peltier signals by varying the magneto-Seebeck coefficient $\Delta S=\Delta S_{AP}-\Delta S_{P}$. Using the results from earlier studies of Walter \textit{et al.} in similar MTJs (Seebeck coefficient values $S_P$= -108 $\mu$V/K and $S_{AP}$= -99 $\mu$V/K), we obtain a Peltier signal $\Delta V=\Delta V_{AP}-\Delta V_{P}$ of  0.37 nV for I = 150 $\mu$A, which is 5 times smaller than our experimental result (1.9 nV, corresponding to 12.5 $\mu \Omega$). Increasing $\Delta S$ by 5 times to 46.2 $\mu$V/K fits our result and gives a $\Delta \Pi$ of 13.4 mV for the magneto-Peltier coefficient of the MTJ.

\subsection{\large {IV. Joule heating in higher order responses}}
In the magneto-Peltier measurements as shown above, non-negligible higher odd harmonic signals have been observed, as plotted in Fig.~\ref{figs2}. In this section, we show that these signals can be ascribed to Joule heating related mechanisms.

\begin{figure}[t]
	\includegraphics[width=13cm]{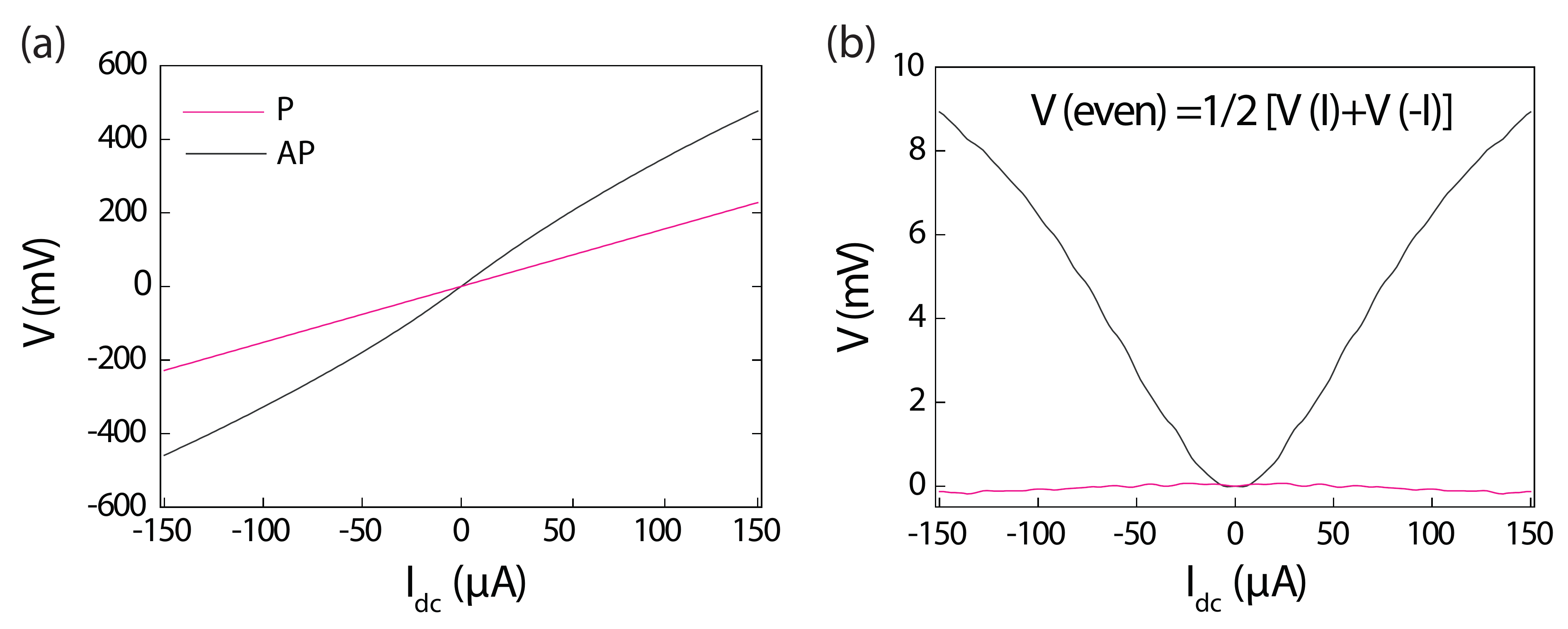}
	\caption{(a) $I-V$ characteristics of TMR for P (pink) and AP (grey) configurations, respectively. (b) The even parts of the $I-V$ characteristics extracted for these two configurations.}
	\label{figs4}
\end{figure}

The resistance of an MTJ is in general bias-dependent, i.e., the $I-V$ curve is nonlinear. The $I-V$ curves we obtain for device A in both the P and AP configurations are shown in Fig.~\ref{figs4}. They are obtained by integrating the differential resistance $dV/dI$ over current, which is acquired using a modulation technique where a small ac current (r.m.s 1 $\mu$A) is superimposed on a dc bias in the lock-in measurements. For the P case, as can be seen from Fig.~\ref{figs4}(a) (pink curve), the $I-V$ characteristic is very close to a linear behavior, which is known as a special property for MgO-based MTJs. In this case the MTJ can be regarded as a normal resistor, where the Joule heating (power dissipation) at the junction ($I\cdot V$) is quadratically dependent on the current. For the AP case, however, the $I-V$ curve is nonlinear especially at higher bias (grey curve), and the Joule heating effects would therefore deviate from a quadratic behavior, showing additional higher order dependences on the current. These additional heating effects thus appear as higher order thermoelectric signals at the thermocouple in the magneto-Peltier measurement.

Suppose the I-V dependence can be expressed as:
\begin{equation}
V(I)=a\cdot I+b\cdot I^2+c\cdot I^3+d\cdot I^4+\ldots
\label{eq:VI}
\end{equation}
where $a$, $b$, $c$ and $d$ are different order response coefficients. The signal at the thermocouple due to Joule heating can then be written as: 
\begin{equation}
V_{thermocouple}(I)= C \cdot(a\cdot I^2+b\cdot I^3+c\cdot I^4+d\cdot I^5+\ldots)
\label{eq:Vthermo}
\end{equation}
where $C$ is a coefficient describing the efficiency of the conversion of heat generated in the MTJ to the voltage at the thermocouple, and is in a unit of A$^{-1}$. Specifically, we are interested in the higher odd order responses shown up in the magneto-Peltier measurement, which come from the even parts of the $I-V$ characteristics (related to coefficients $b$, $d$, \ldots), as can be directly seen from Eq.~\eqref{eq:VI} and Eq.~\eqref{eq:Vthermo} . To discuss them separately, we extract the even parts of the $I-V$ curves, by performing
\begin{equation}
V (\textup{even})=\frac{1}{2}[V(+I)+V(-I)]    
\label{eq:Veven}
\end{equation}
on the $I-V$ curves, and the results are shown in Fig.~\ref{figs4}(b). Note that in Ref. \cite{zhang_seebeck_2012}, this part of signal is regarded as magneto-Seebeck signal as a result of the asymmetric heating induced temperature gradient across the MTJ. Here, however, using our finite element model we find this signal to be too large if explained in that way. We thus interpret the even parts as the intrinsic $I-V$ characteristics of the MTJ, due to the inevitable difference between the two interfaces across MgO.

For P configuration, the $I-V$ curve follows an almost-linear behavior, and the even part is much smaller compared to the AP configuration. The higher harmonic signals generated from P configuration are all around 0 nV. From here, we only discuss the AP case.

Now we calculate the odd Joule heating signal expected at thermocouple that is caused by this even part, by multiplying the coefficient $C$ defined in Eq.~\eqref{eq:Vthermo}. $C$ can be determined in the much simpler parallel case, by comparing the overall signal on thermocouple which is dominated by Joule heating with the $I-V$ curve. Multiplying $C$ and $I$ with the grey curve in Fig.~\ref{figs4}(b), we obtain an expected signal as the red solid line shown in Fig. S6. This is the expected signal on thermocouple that originates from the even $I-V$ characteristics for the AP configuration due to dissipation. 

\begin{figure}[b]
	\includegraphics[width=7cm]{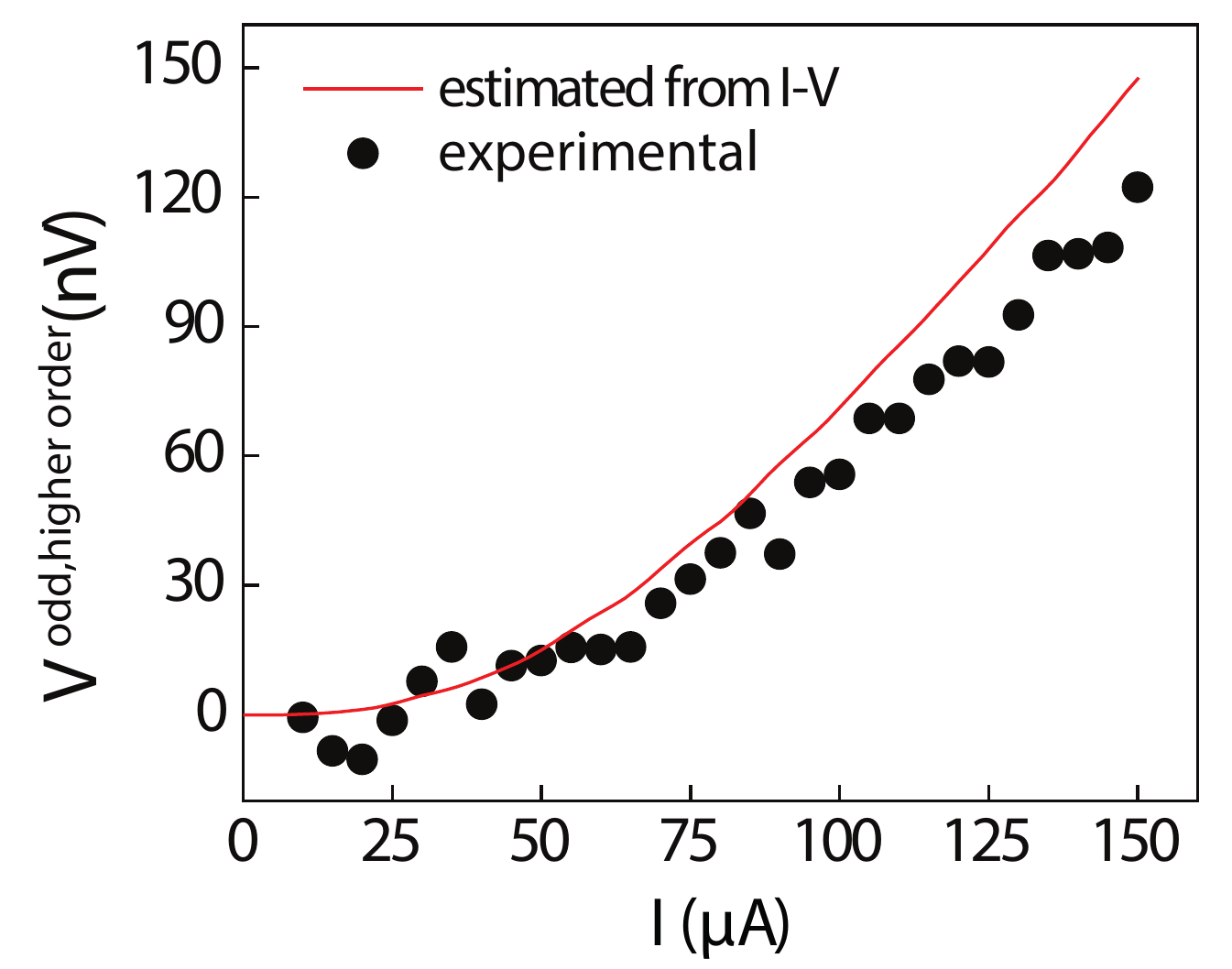}
	\caption{Odd Joule heating signal that is expected to show on thermocouple in magneto-Peltier measurement configuration (red curve) and odd responses from magneto-Peltier measurement configuration that are calculated from lock-in measurements (black dots). Only the positive bias is shown.}
	\label{figs5}
\end{figure}

On the other hand, the odd order responses can be obtained by performing
\begin{equation}
\begin{aligned}
V^{\textup{odd\ order\ responses}} {} & =R_1I_0+R_3I_0^3+R_5I_0^5\ (+R_7 I_0^7) \\ & =V^{1f}+V^{3f}-V^{5f}\ (-V^{7f}) 
\end{aligned}
\label{eq:Vodd}
\end{equation}
on the signals we obtained from lock-in measurements. The $V^{7f}$ is ignored here, as it is too small compared to other signals therefore omitting it does not change the overall analysis. Subtracting the linear response which we already know that originates from Peltier effect, we can get the odd non-linear signals, as plotted in Fig.~\ref{figs5} as black dots. It can be seen that the measured signals are very close to the expected values. It suggests that the nonlinear thermoelectric signals are indeed from Joule heating.

In summary, the even component of the TMR $I-V$ curve, that we clearly observe in our measurements, can produce Joule heating that shows higher odd order dependence with the current which mimics the magneto-Peltier effect. Note that this implies that Joule heating is different with opposite tunneling current direction, a similar observation as in Ref. \cite{gapihan_heating_2012}, though the interpretation we have (based on careful harmonics analysis) is different from theirs (dissipation occurs at different contacts for opposite current direction). By comparing with our lock-in measurement signals detected on thermocouple we conclude that the higher odd harmonic signals that we observed are generated by this higher-order Joule heating effect. 

\end{document}